\documentclass[11pt,twoside]{amsart}
\usepackage{amssymb,latexsym}
\usepackage{epic}
\usepackage[a4paper, margin=2cm]{geometry}

\usepackage{pifont,pxfonts,txfonts}
\usepackage{yfonts}
\usepackage{bbm}
\usepackage{calrsfs}
\usepackage{empheq}
\usepackage{color}
\usepackage[colorlinks,linkcolor=red,anchorcolor=blue,citecolor=green]{hyperref}
\usepackage{amsmath, amstext, amsbsy, amscd}
\usepackage[mathscr]{eucal}

\usepackage{times}

  \usepackage{amsmath, amsthm, amsfonts,extarrows}

\newtheorem{theorem}{Theorem}[section]

   \newcommand{\ba}{\begin{eqnarray}}
   \newcommand{\na}{\end{eqnarray}}
   \newcommand{\ban}{\begin{eqnarray*}}
   \newcommand{\nan}{\end{eqnarray*}}

\begin{document}

\title[Quantum McKay for Disc Invariants]{Quantum McKay Correspondence for Disc Invariants of Toric Calabi-Yau $3$-orbifolds}

\author[Hua-Zhong Ke]{Hua-Zhong Ke}
  \address{Department of Mathematical Sciences\\ Tsinghua University\\ Beijing, 100084\\ China}
  \email{kehuazh@mail.sysu.edu.cn}

\author[Jian Zou]{Jian Zhou}     
    \address{Department of Mathematical Sciences\\ Tsinghua University\\ Beijing, 100084\\ China}
    \email{jzhou@math.tsinghua.edu.cn}

\maketitle%

\begin{abstract}
We announce a result on quantum McKay correspondence for disc invariants of outer legs in toric Calabi-Yau $3$-orbifolds, and illustrate our method in a special example $[\mathbb C^3/\mathbb Z_5(1,1,3)]$. 

%\ \\
%Keywords: Quantum McKay correspondence, Disc invariants, Open mirror symmetry     

%\ \\
%MRSubClass: 14N35
\end{abstract}

\section{Introduction}

Let $\hat{\mathcal X}\rightarrow\mathcal X$ be a crepant resolution of a Gorenstein orbifold $\mathcal X$. McKay correspondence predicts that the invariants of $\hat{\mathcal X}$ coincide with the orbifold invariants of $\mathcal X$. See \cite{Reid} for an exposition of such results for classical invariants. Physicists and mathematicians are also interested in quantum version of McKay correspondence, which concerns about quantum invariants, e.g. Gromov-Witten invariants. A famous example is the Crepant Resolution Conjecture (CRC) proposed by Ruan \cite{Ruan}, which relates the orbifold small quantum cohomology of $\mathcal X$ to the small quantum cohomology of $\hat{\mathcal X}$ in a subtle way. See \cite{BG, CCIT, H, Z08} and references therein for refinements and recent results.

Our point of view towards quantum McKay correspondence is to understand both the toric Calabi-Yau orbifold and its toric crepant resolution as symplectic reductions of the same system of charge vectors. Quantum McKay correspondence is related to phase change phenomena in string theory. Roughly speaking, physicists expect that string theories at different limit points in some (complexified) K\"ahler moduli spaces should be related to each other. In the early 90's, Witten proposed gauged linear sigma model (GLSM) to study phase change phenomena \cite{W}. Mathematically, GLSM realizes the target space of Gromov-Witten theory as a symplectic reduction of a complex vector space by some linear action by a torus. The weights of the torus action are some vectors with integral components. They are called charge vectors in the physics literature. For different regular values of the moment map of the torus action, the resulting symplectic reductions are related by birational transformations called flops. It is generally expected that their Gromov-Witten invariants are related to each other in a suitable way. In performing the symplectic reduction, one can also produce orbifolds instead of manifolds. In this way, one can put the toric Calabi-Yau orbifold and its toric crepant resolutions in the same family.

For a toric Calabi-Yau $3$-fold obtained in this way, one can use its charge vectors to compute genus-zero (closed) Gromov-Witten invariants \cite{CKYZ} and disc invariants \cite{AV,AKV,LM}. Similarly, it can also be done for toric Calabi-Yau $3$-orbifolds. In this note, we will consider disc invariants (with respect to Aganagic-Vafa branes), which are open Gromov-Witten invariants. In \cite{KZ12, Z11}, we observed that superpotentials of disc invariants for different limit points in the extended (complexified) K\"ahler moduli of $[\mathbb C^3/G]$ together with an Aganagic-Vafa brane are governed by different sets of charge vectors which span the same linear subspace, and in the effective case, the superpotentials are equal to each other as formal power series, via the change of variables which can be read off directly from the charge vectors. This matches with the prediction by physicists \cite{BKMP09,BKMP10} and provides a natural derivation of the mysterious change of variables proposed in \cite{BKMP09,BKMP10}. In a forthcoming paper \cite{KZ13}, we will generalize our result to all toric Calabi-Yau $3$-orbifolds, using the open mirror symmetry recently proved in \cite{FLT}.

For recent related results of quantum McKay correspondence of disc invariants, see \cite{BC,BCR,CR}. See also \cite{CCLT12} for disc invariants of FOOO type. Recently, Chan, Cho, Lau and Tseng have proved CRC of disc invariants of FOOO type for all semi-projective toric CY orbifolds with canonical Gorenstein singularities \cite{CCLT13}. Note that FOOO type invariants considered by these authors and the invariants for Aganagic-Vafa branes considered by us involve Lagrangian submanifolds of different Maslov indices.

The rest of the paper is organized as follows. In Section 2, to illustrate our method, we present some details of the proof of the quantum McKay correspondence of disc invariants for $[\mathbb C^3/\mathbb Z_5(1,1,3)]$. In Section 3, we announce the quantum McKay correspondence for disc invariants of effective outer legs in toric Calabi-Yau $3$-orbifolds.

\section{A special case: $[\mathbb C^3/\mathbb Z_5(1,1,3)]$}

In this section, let $\mathcal X=[\mathbb C^3/\mathbb Z_5(1,1,3)]$ and let $\hat{\mathcal X}$ be the unique toric crepant resolution of $\mathcal X$. By $\mathbb Z_5(1,1,3)$, we mean the subgroup of $SL(3,\mathbb C)$ generated by diag$(e^{2\pi i/5}, e^{2\pi i/5}, e^{3\cdot 2\pi i/5})$. We will establish the quantum McKay correspondence for disc invariants with Aganagic-Vafa branes for this case. We will achieve this in the following steps:
\begin{itemize}
\item[Step 1]Using open mirror symmetry to compute the superpotential functions of Aganagic-Vafa disc invariants for $\mathcal X$ and $\hat{\mathcal X}$.
\item[Step 2]Find the corresponding charge vectors for $\mathcal X$ and $\hat{\mathcal X}$ from their superpotential functions.
\item[Step 3]Find the formula for change of variable from the change of charge vectors.
\item[Step 4]Identify the superpotential functions after the change of variables.
\end{itemize}
The same steps can be carried out for general toric Calabi-Yau $3$-orbifolds.

We refer readers to the book \cite{CLS} for basic facts of toric geometry used in this section. The toric crepant resolution  $\hat{\mathcal X}\rightarrow\mathcal X$ is given by Figure 1,
\begin{figure}
\begin{center}
\setlength{\unitlength}{1cm}
\begin{picture}(0,4)(2,-2)
\drawline(0,0)(-2,-2)(0,1)(1,0)(-1,-1)(0,1)(0,0)(1,0)(-2,-2)
\put(0,0){\circle*{0.1}}
\put(-1,-1){\circle*{0.1}}
\put(0,1){\circle*{0.1}}
\put(1,0){\circle*{0.1}}
\put(2,0){\vector(1,0){1}}
\put(-2,-2){\circle*{0.1}}

\drawline(2,-2)(4,1)(5,0)(2,-2)
\put(4,0){\circle*{0.1}}
\put(3,-1){\circle*{0.1}}
\put(2,-2){\circle*{0.1}}
\put(4,1){\circle*{0.1}}
\put(5,0){\circle*{0.1}}
\put(5,0.1){$b_2$}
\put(4,1.1){$b_3$}
\put(3,-0.9){$b_5$}
\put(4,0.1){$b_4$}
\put(1.6,-2){$b_1$}
\end{picture}
\end{center}
\caption{Toric resolution of $\mathcal X=[\mathbb C^3/\mathbb Z_5(1,1,3)]$}
\end{figure}
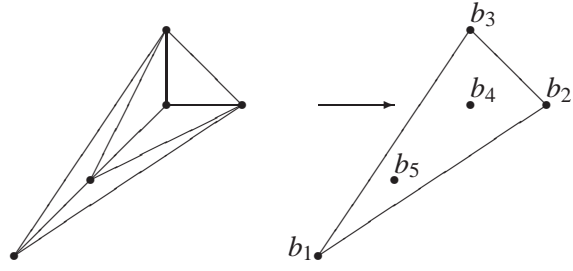
where $$b_1=\left[\begin{array}{c}-2\\-2\\1\end{array}\right],b_2=\left[\begin{array}{c}1\\0\\1\end{array}\right],b_3=\left[\begin{array}{c}0\\1\\1\end{array}\right],b_4=\left[\begin{array}{c}0\\0\\1\end{array}\right],b_5=\left[\begin{array}{c}-1\\-1\\1\end{array}\right].$$

More explicitly, choose two charge vectors $(1,0,0,1,-2),(0,1,1,-3,1)$ for $\hat{\mathcal X}$, and let $(\mathbb C^*)^2$ act on $\mathbb C^5=\textrm{Spec}\mathbb C[Z_1,\dots,Z_5]$ by $$(u_1,u_2).(Z_1,\cdots,Z_5)=(u_1Z_1,u_2Z_2,u_2Z_3,u_1u_2^{-3}Z_4,u_1^{-2}u_2Z_5).$$ Then $$\mathcal X=[\mathbb C^3\times(\mathbb C^*)^2/(\mathbb C^*)^2],$$ and $$\hat{\mathcal X}=[\mathbb C^5\setminus\mathbb V(Z_1Z_2,Z_1Z_3,Z_1Z_5,Z_2Z_4,Z_3Z_4)/(\mathbb C^*)^2].$$

From the above chosen charge vectors, the secondary fan is given by Figure 2:
\begin{figure}
\begin{center}
\setlength{\unitlength}{1cm}
\begin{picture}(0,6)(0,-3)
\put(0,0){\vector(1,0){1}}
\put(0,0){\vector(0,1){1}}
\put(0,0){\vector(-2,1){2}}
\put(0,0){\vector(1,-3){1}}

\drawline(0.5,0.5)(1.5,1.5)(1.5,2)(2,1.5)(1.5,1.5)
\drawline(0.5,0.5)(1.5,2)
\drawline(0.5,0.5)(2,1.5)
\drawline(1.5,2)(1,1)(2,1.5)
\put(0.5,0.5){\circle*{0.1}}
\put(1,1){\circle*{0.1}}
\put(1.5,1.5){\circle*{0.1}}
\put(1.5,2){\circle*{0.1}}
\put(2,1.5){\circle*{0.1}}

\drawline(1,-2)(1.5,-1.5)(2,-0.5)(2.5,-1)(1.5,-1.5)
\drawline(1,-2)(2,-0.5)
\drawline(1,-2)(2.5,-1)
\put(1,-2){\circle*{0.1}}
\put(1.5,-1.5){\circle*{0.1}}
\put(2,-0.5){\circle*{0.1}}
\put(2.5,-1){\circle*{0.1}}
\put(2,-1){\circle*{0.1}}

\drawline(-1.5,-2)(0,-1)(-0.5,-0.5)(-1.5,-2)
\put(-1.5,-2){\circle*{0.1}}
\put(-1,-1.5){\circle*{0.1}}
\put(-0.5,-0.5){\circle*{0.1}}
\put(0,-1){\circle*{0.1}}
\put(-0.5,-1){\circle*{0.1}}

\drawline(-1.5,1)(0,2)(-0.5,2.5)(-1.5,1)(-0.5,2)(-0.5,2.5)(0,2)(-0.5,2)
\put(-1.5,1){\circle*{0.1}}
\put(-1,1.5){\circle*{0.1}}
\put(-0.5,2.5){\circle*{0.1}}
\put(0,2){\circle*{0.1}}
\put(-0.5,2){\circle*{0.1}}

\put(1,0){$(1,0)$}
\put(-0.3,1.1){$(0,1)$}
\put(-3.3,1){$(-2,1)$}
\put(1,-3){$(1,-3)$}
\end{picture}
\end{center}
\caption{Secondary fan for $\mathcal X=[\mathbb C^3/\mathbb Z_5(1,1,3)]$}
\end{figure}
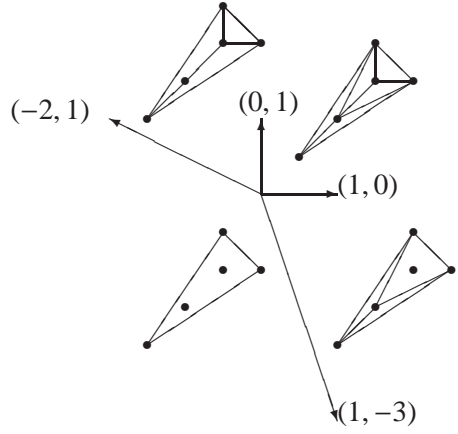
For (semi-projetive) toric Calabi-Yau $3$-orbifolds, we have a working definition for open Gromov-Witten invariants \cite{FLT} via virtual localization \cite{GP}, and we can use topological vertex to calculate them \cite{Ross}. Topological vertex is very effective in the manifold case \cite{AKMV,LLLZ}, but not that effective in the orbifold case since it involves Hurwitz-Hodge integrals which are difficult to compute. Thus in general, it is difficult to compute the open Gromov-Witten invariants. The open mirror symmetry \cite{FL,FLT} expresses the disc potential, the generating function of disc invariants, in terms of the superpotential, a generalized hypergeometric series, via open-closed mirror map. 

Let $\mathcal L$ be an Aganagic-Vafa brane in $\mathcal X$, intersecting the toric curve given by $b_2$ and $b_3$. By \cite{FLT}, the disc potential of $(\mathcal X,\mathcal L)$ with framing $f$ is equal to the following superpotential via open-closed mirror map:
\begin{eqnarray*}
W_f^{\mathcal X,\mathcal L}(q_4,q_5,q_0)&=&\sum\limits_{\substack{m_0\in\mathbb Z_{>0},m_4,m_5\in\mathbb Z_{\geqslant 0},\\\frac{m_0}{5}-\frac{m_4}{5}-\frac{3}{5}m_5\in\mathbb Z_{\geqslant 0}}}\frac{(-1)^{\lfloor-\frac{2}{5}m_4-\frac{1}{5}m_5\rfloor+m_0(1-\frac{f+1}{5})}q_0^{m_0}q_4^{m_4}q_5^{m_5}}{\frac{m_0}{5}\Gamma(1+\frac{m_0}{5}-\frac{m_4}{5}-\frac{3}{5}m_5)\Gamma(1+m_4)\Gamma(1+m_5)}\\
&&\quad\quad\cdot\frac{\Gamma(\frac{f+1}{5}m_0+\frac{2}{5}m_4+\frac{m_5}{5})}{\Gamma(1+\frac{f}{5}m_0-\frac{2}{5}m_4-\frac{m_5}{5})}.
\end{eqnarray*}

On the other hand, let $\hat{\mathcal L}$ be an Aganagic-Vafa brane in $\hat{\mathcal X}$, intersecting the toric curve given by $b_2$ and $b_3$. By \cite{FLT}, the disc potential of $(\hat{\mathcal X},\hat{\mathcal L})$ with framing $\hat f$ is equal to the following superpotential via open-closed mirror map:
\begin{eqnarray*}
W_{\hat f}^{\hat{\mathcal X},\hat{\mathcal L}}(\hat q_1,\hat q_5,\hat q_0)&=&\sum\limits_{\substack{\hat m_0\in\mathbb Z_{>0},\hat m_1,m_5\in\mathbb Z_{\geqslant 0},\\\hat m_0-5\hat m_1-3\hat m_5\in\mathbb Z_{\geqslant 0}}}\frac{(-1)^{\hat 2m_1+\hat m_5+(\hat f+\frac{2}{5})\hat m_0+\lfloor-\frac{2}{5}\hat m_0\rfloor}\hat q_0^{\hat m_0}\hat q_1^{\hat m_1}\hat q_5^{\hat m_5}}{\hat m_0\Gamma(1+\hat m_0-5\hat m_1-3\hat m_5)\Gamma(1+\hat m_1)\Gamma(1+\hat m_5)}\\
&&\quad\quad\cdot\frac{\Gamma((\hat f+1)\hat m_0-2\hat m_1-\hat m_5)}{\Gamma(1+\hat f\hat m_0+2\hat m_1+\hat m_5)}.
\end{eqnarray*}

According to \cite{KZ12,KZ13}, one can read off the charge vectors for $(\mathcal X,\mathcal L)$ and $(\hat{\mathcal X},\hat{\mathcal L})$ from the superpotentials, and the result is as follows. The extended charge vectors for $(\mathcal X,\mathcal L)$ are
$$\begin{array}{ccccccccc}
l^{(4)}&=&(-\frac{1}{5},&-\frac{2}{5},&-\frac{2}{5},&1,&0,&0,&0),\\
l^{(5)}&=&(-\frac{3}{5},&-\frac{1}{5},&-\frac{1}{5},&0,&1,&0,&0),\\
l^{(0)}&=&(\frac{1}{5},&\frac{f}{5},&-\frac{f+1}{5},&0,&0,&1,&-1),
\end{array}$$
and the extended charge vectors for $(\hat{\mathcal X},\hat{\mathcal L})$ are
$$\begin{array}{ccccccccc}
\hat l^{(1)}&=&(1,&2,&2,&-5,&0,&0,&0),\\
\hat l^{(5)}&=&(0,&1,&1,&-3,&1,&0,&0),\\
\hat l^{(0)}&=&(0,&\hat f,&-\hat f-1,&1,&0,&1,&-1),
\end{array}$$
By comparing these charge vectors, one notices that $\{l^{(4)},l^{(5)},l^{(0)}\}$ and $\{\hat l^{(1)},\hat l^{(5)},\hat l^{(0)}\}$ span the same linear subspace of $\mathbb Q^5$ if and only if $f=5\hat f+2$, and under this condition, the charge vectors are linear related to each other as follows:
\begin{eqnarray*}
\left\{\begin{array}{ccc}l^{(4)}&=&-\frac{1}{5}\hat l^{(1)},\\l^{(5)}&=&-\frac{3}{5}\hat l^{(1)}+\hat l^{(5)},\\l^{(0)}&=&\frac{1}{5}\hat l^{(1)}+\hat l^{(0)}.\end{array}\right.
\end{eqnarray*}
These linear relations give us the following formula for change of variables when $f=5\hat f+2:$
\begin{eqnarray*}
\left\{\begin{array}{ccc}q_4&=&\hat q_1^{-\frac{1}{5}},\\q_5&=&\hat q_1^{-\frac{3}{5}}\hat q_5,\\q_0&=&\hat q_1^{\frac{1}{5}}\hat q_0.\end{array}\right.
\end{eqnarray*}
Now we check directly that, under the above change of variables, we have
\begin{eqnarray*}
W_{f(\hat f)}^{\mathcal X,\mathcal L}(q(\hat q),q_0(\hat q,\hat q_0))&=&5\sum\limits_{\substack{\hat m_0\in\mathbb Z_{>0},\hat m_1,m_5\in\mathbb Z_{\geqslant 0},\\\hat m_0-5\hat m_1-3\hat m_5\in\mathbb Z_{\geqslant 0}}}\frac{(-1)^{\hat 2m_1+\hat m_5+(\hat f+\frac{2}{5})\hat m_0}\hat q_0^{\hat m_0}\hat q_1^{\hat m_1}\hat q_5^{\hat m_5}}{\hat m_0\Gamma(1+\hat m_0-5\hat m_1-3\hat m_5)\Gamma(1+\hat m_1)\Gamma(1+\hat m_5)}\\
&&\quad\quad\cdot\frac{\Gamma((\hat f+1)\hat m_0-2\hat m_1-\hat m_5)}{\Gamma(1+\hat f\hat m_0+2\hat m_1+\hat m_5)}\cdot(-1)^{\lfloor-\frac{2}{5}\hat m_0\rfloor}\\
&=&5W_{\hat f}^{\hat{\mathcal X},\hat{\mathcal L}}(\hat q,\hat q_0).\end{eqnarray*}
This establishes the correspondence of disc potentials of $(\mathcal X,\mathcal L)$ and $(\hat{\mathcal X},\hat{\mathcal L})$.

\section{General case}

In this section, we state our result of quantum McKay correspondence for disc invariants. This can be obtained by following the same step as in the last section. The details of the proof will appear in a forthcoming paper \cite{KZ13}.

Let $\hat{\mathcal X}\xrightarrow{\pi}\mathcal X$ be a crepant resolution of the Calabi-Yau $3$-orbifold $\mathcal X$. Assume that $\mathcal L$ is an outer Aganagic-Vafa brane in $\mathcal X$, intersecting an effective noncompact toric curve $\mathcal C$ in $\mathcal X$.  By the open mirror symmetry recently proved in \cite{FLT}, the disc potential of $(\mathcal X,\mathcal L)$ with framing $f$ is equal to the the superpotential $W^{\mathcal X,\mathcal L}_f(q,q_0)$. Note that $\pi^{-1}(\mathcal C)$ is also a noncompact toric curve in $\hat{\mathcal X}$, and we assume that $\hat{\mathcal L}$ is an outer Aganagic-Vafa brane in $\hat{\mathcal X}$ intersecting $\pi^{-1}(\mathcal C)$. Again by the open mirror symmetry, the disc potential of $(\hat{\mathcal X},\hat{\mathcal L})$ with framing $\hat f$ is equal to the the superpotential $W^{\hat{\mathcal X},\hat{\mathcal L}}_{\hat f}(\hat q,\hat q_0)$. One can use the superpotential functions in both cases to read off the charge vectors and hence make the corresponding change of variables to identify the superpotential functions.

\begin{theorem}
We have $$W^{\mathcal X,\mathcal L}_{f(\hat f)}(q(\hat q),q_0(\hat q_0))=s_1W^{\hat{\mathcal X},\hat{\mathcal L}}_{\hat f}(\hat q,\hat q_0),$$ where $s_1$ is an integer determined by the toric data, and the change of variables $f=f(\hat f),q=q(\hat q),q_0=q_0(\hat q_0)$ can be read off directly from the charge vectors.
\end{theorem}

This result establishes the correspondence of disc potentials of $(\mathcal X,\mathcal L)$ and $(\hat{\mathcal X},\hat{\mathcal L})$. We also have a similar result for ineffective outer legs.

{\bf Acknowledgements.} 
The first author would like to thank Renzo Cavalieri and Mark Shoemaker for helpful discussions, and Yongbin Ruan for his interest in this work. We thank the referees for their time and helpful comments.The first author is partially supported by China Scholarship Council, and the second author is partially supported by NSFC grant 11171174.

\end{document}